\documentclass[]{aastex631}

\begin{document}

\title{Proto-Planetary Nebulae Are Not the Precursors of Planetary Nebulae With Close Binary Central Stars}

%\correspondingauthor{August Muench}
%\email{greg.schwarz@aas.org, gus.muench@aas.org}

\author[0000-0002-0816-1090]{Todd C. Hillwig}
\affiliation{Dept. of Physics \& Astronomy, Valparaiso University, Valparaiso, IN 46383 USA}

\author[0000-0003-1471-8892]{Bruce J. Hrivnak}
\affiliation{Dept. of Physics \& Astronomy, Valparaiso University, Valparaiso, IN 46383 USA}

\author[0000-0003-3947-5946]{David Jones}
\affiliation{Instituto de Astrof\'isica de Canarias, E-38205 La Laguna, Tenerife, Spain}
\affiliation{Departamento de Astrof\'isica, Universidad de La Laguna, E-38206 La Laguna, Tenerife, Spain}

%\author[0000-0002-1126-869X]{Orsola De Marco}
%\affiliation{Macquarie University}

\correspondingauthor{Todd C. Hillwig}


\begin{abstract}

Proto-planetary nebulae are considered to be the evolutionary link between AGB stars and planetary nebulae. They possess many of the physical characteristics expected of such a link and provide valuable information about the evolutionary process in these objects. One important characteristic of proto-planetary nebulae is the prevalence of very strong bipolar and point-symmetric morphologies. In planetary nebulae, these shapes are widely considered to be the result of shaping by a binary companion. Indeed a number of studies have shown links between strong axial or point-symmetry in planetary nebulae and {\it close} binary central star systems that have gone through a common envelope phase. Here, we explore whether the observed proto-planetary nebulae may be the progenitors of planetary nebulae with close binary central stars. We find that despite the strong shaping of proto-planetary nebulae this is unlikely to be the case. We also discuss options for evolution to close binaries and the alternatives.

\end{abstract}

\keywords{}


\section{Introduction} \label{sec:intro}

The observational discovery of proto-planetary nebulae (PPNe)  began following the IRAS sky survey, which identified a sample of objects with strong mid-infrared excesses due to cool circumstellar dust \citep{parthasarathy1986}. This opened up new avenues for the study of the late stages of evolution for intermediate mass stars. The question of how essentially spherical atmospheres of asymptotic giant branch (AGB) stars can evolve to become planetary nebulae (PNe) which show a panoply of morphologies \citep[e.g.,][]{zhang1998} has been a subject of debate for some time. The availability of objects in transition between these stages provided an exciting opportunity to answer many existing questions, especially in regards to the mechanisms that shape the expanding AGB envelope. One question--when does the shaping take place--seemed to be answered quickly; {\it very} early in the evolution. This was a result of the detection, via Hubble Space Telescope (HST) imagery, of strongly bipolar and multi-polar PPNe with central stars of F--G spectral type. Early shaping provides limits on the interacting stellar winds (ISW) and generalized interacting stellar winds (GISW) models \citep{kwok1978,balick1987}, meaning that most of the shaping needed to be performed as part of the ejection of the envelope rather than through gradual shaping of density enhancements by a follow-up fast wind. One method of producing such structure during ejection is through common envelope (CE) evolution with a binary companion \citep{nordhaus06}.

There is strong observational evidence that CE evolution is responsible for producing close binary central stars of PNe \citep[CSPNe,][]{hillwig16}. In this case, CE evolution is the primary mechanism for moving the post-AGB core close to a stellar companion, typically with an ending orbital period of $\lesssim 2$ days. There is also a strong connection between close binary stars and the shaping of PNe \citep[e.g.,][]{hillwig16,miszalski09b}.

Since CE evolution is a way to shape the ejected nebula during the PPN phase and since it is also the mechanism thought responsible for the closeness of the binary pair, the question naturally arises: are PPNe, especially those showing strong axisymmetries, the progenitors of close binary CSPNe? We begin by reviewing our current knowledge of PPNe, CE evolution, and close binary CSPNe. We then address whether the results suggest, or allow, a relationship between PPNe and close binary CSPNe.

\section{A Brief Review of Proto-Planetary Nebulae} \label{sec:PPNe}

PPNe candidates were first identified by \citet{parthasarathy1986} who associated several bright cataloged F supergiant stars with IRAS dust sources, classifying them as post-AGB stars. Likewise, \citet{hrivnak1988} used an IRAS identification to discover an optically associated bright star with an F supergiant spectrum, specifically classifying the object as a PPN. \citet{hrivnak1989} followed this up with a further identification of eight similar objects. A multiwavelength study by \citet{vanderveen1989}  identified 42 candidate objects in this transitional phase.

Resolved images of PPNe were first provided using HST/WFPC2 for individual PPNe by \citet[IRAS~17150-3224]{kwok1998} and \citet[IRAS~1741-2411]{su1998}, and a large survey of 27 objects by \citet{ueta2000}. 21 of these nebulae were found to be either bipolar or clearly elongated and 6 were point-like. Later surveys by \citet{sahai2007} and \citet{siodmiak2008} also show either elongated or bipolar nebulosity, or no nebulosity (stellar profiles).

A further study of PPN morphologies, along with AGB nebulae and young PNe, is provided in an online catalog by Balick and collaborators\footnote{http://faculty.washington.edu/balick/pPNe/}. They also find that the nebulosity in nearly all resolved PPNe is elongated or bipolar, with very few round PPNe. However, they do suggest that round PPNe may be faint and difficult to disentangle from the Airy rings of the bright central star. Many of the bipolar PPNe show two main morphological components: a torus and jets (also called high velocity outflows). Given the preponderance of bipolar and multi-polar, along with clearly elongated elliptical, morphologies, for the remainder of the paper we will focus on these objects. Mildly elliptical or round PPNe may have different properties or may simply be viewed at different inclinations, but at present there is not enough observational data to describe those separately and to include them in this work.

\subsection{Ages of PPNe}
A number of studies have used distances, angular sizes, proper motions, and radial velocity measurements from spectra to determine ages of individual PPNe. 
See \citet{balick2013} for an example of the proper motion and angular size method and \citet{velazquez2014} for morphokinematical modeling using spatially resolved radial velocities and geometrical models, both for CRL~618.
In these cases, age refers to time since ejection of the visible nebulosity. Table \ref{tab:ppn_ages} lists 9 PPNe for which ages have been determined. It is possible that a few of the objects in Table \ref{tab:ppn_ages} may be misclassified young PNe, where we use the standard definition for transition from PPN to PN as when $T_\mathrm{eff}$ of the CS increases to roughly 30,000 K, at which point it is able to photoionize the surrounding nebula (though the observational classification can be difficult to determine precisely). Column two of the table lists ages from the original studies. These are based on expansion ages and use the distances given in column three. Updated distances from Gaia DR3, where available, are given in column four with the corresponding RUWE values in column five \citep{bailer-jones2021}. Updated ages based on Gaia distances are provided in column six. The calculated expansion ages range from a low of ~100 yr to a high of ~4000 yr (ignoring the updated distance for M 2-9 whose Gaia parallax has a very high RUWE value and thus is very uncertain). However, only two of the 9 have ages $\gtrsim2000$ yr. All of these objects are clearly elongated and show at least some bipolar or multi-polar morphology. Whether elliptical PPNe have similar ages is not currently known, though some of these objects show an elliptical component along with bipolar outflows and in each case the elliptical component's age seems consistent with the bipolar component. Also, \citet{garcia2022} demonstrate that the CE phase can produce elliptical nebulae (in fact, their study results in more elliptical than bipolar nebulae) and that they have the same age ranges as bipolar nebulae.

\citet{huggins2007} provided ages for most of the PPNe in Table \ref{tab:ppn_ages}, using angular sizes and proper motion values to determine those ages. \citet{huggins2007} also demonstrated that the tori and jets/lobes were formed at measurably different times, with the jets/lobes typically a few hundred years younger than the tori. This result is supported in strongly bipolar PNe by \citet{moragabaez2026} who show dynamical ages of tori that are larger than those of their bipolar lobes in five of the six objects studied. \citet{moragabaez2026} used the radial velocity of molecular gas along with angular sizes and an assumption that the torii are circular. The circular assumption is well supported in their results by finding torus inclinations consistent with independent measures of inclination for bipolar lobes and other structures from previous studies. The studies cited in Table \ref{tab:ppn_ages} for PPNe overwhelmingly show that the ejection of a given component occurs over a short time such that it can be treated as nearly simultaneous ejection.  

\subsection{Energetics and Shaping of PPNe\label{PPN shaping}}
\citet{bujarrabal2001,bujarrabal2004} measured the kinetic energies and linear momenta of PPNe and showed that in at least $\sim 80$\% they are too large to be due to radiation pressure. \citet{huggins2007} also suggested that the similar component ages are consistent with formation from CE evolution. \citet{sahai2017} in a study of the PPN IRAS~16342-3814 show that either Roche lobe overflow from a primary onto a companion, or accretion within a CE are most energetically likely to create the observed collimated outflows. And \citet{nordhaus2007} show that binary interactions, particularly through CE evolution, provide the most robust physical processes for the formation of bipolar PPNe and their associated jets.
Papers by \citet{garcia2021,garcia2022} and \citet{zou2020} use CE evolution simulations to evolve post-CE systems from PPNe to PNe and show that CE evolution can produce both bipolar and elliptical PNe through the PPN phase, and with similar ages, as mentioned above. In addition, \citet{blackman2014} conclude that energetically most PPNe require a CE interaction. Overall, studies point to a CE origin as an energetically likely way to produce the strongly shaped morphologies seen in PPNe.

%Table 1 - PPN Ages
\begin{table}
	\centering
	\caption{Properties of PPNe from the literature along with updated ages based on Gaia distances.}
	\label{tab:ppn_ages}
	\begin{tabular}{lccccccc} % 
    	\hline
		PPN name & Published Age\tablenotemark{a} & Dist.\tablenotemark{a} &  Gaia Dist. & RUWE & Updated Age\tablenotemark{b} & $\log T_\mathrm{eff}$\tablenotemark{a}  & Lum\tablenotemark{a} \\
                & (yr)           & (kpc) & (kpc) &   & (yr) &   & L$_\odot$ \\
		\hline
        CRL~618                 & $100^1$, $100^2$, $422^3$   & $0.9^3$      & ---  & ---  & --- & --- & --- \\
        IRAS~16342-3814         & $160^4$             & $2.0^4$      & 3.06 & 1.04 & 245 & --- & --- \\
        Hen~3-1475              & $878^3$             & $5.8^3$      & 5.32 & 1.66 & 805 & $4.1^{10}$ & $(11000^{+11000}_{-5000})^{10}$ \\
        M~1-92                  & $900^5$, $1520^3$       & $2.5^\mathrm{3,5}$      & 1.57 & 1.61 & 500, 845 & $4.49^{10}$ & $(2160^{+680}_{-480})^{10}$\\
        M~2-56                  & 1000-$1700^6$, $1750^3$ & $3.0^6$, $2.1^3$ & ---  & ---  & --- & --- & $6000\pm1500^{11}$ \\
        M~2-9\tablenotemark{c}  & $1300^3$            & $0.64^3$     & 3.24 & 6.79 & 6580\tablenotemark{d}& $3.90^{12}$ & $2000\pm750^{11}$ \\
        IRAS~19475+3119         & $1900^7$            & $4.8^7$      & 2.97 & 1.43 & 1175 & $3.83^{10}$ & $(22300^{+2500}_{-2700})^{10}$\\
        M~1-16\tablenotemark{c} & $2000^3$,$8700^8$       & $1.8^3$,$6.2^8$      & ---  & ---  & --- & --- & ---\\
        IRAS~19024+0044         & $2870^9$            & $3.5^9$      & 5.21 & 1.26 & 4272 & --- & $12000\pm3000^{11}$\\
		\hline
	\end{tabular}
 \tablenotetext{a}{References in this table: 1) \citet{balick2013}, 2) \citet{velazquez2014}, 3) \citet{huggins2007}, 4) \citet{sahai2017}, 5) \citet{bujarrabal1998}, 6) \citet{castro2002},7) \citet{sanchez2006}, 8) \citet{gomez-munoz2023}, 9) \citet{sahai2005}, 10) \citet{oudmaijer2022}, 11) \citet{vickers2015}, 12) Gaia DR3}
 \tablenotemark{b}{Using distances based on Gaia DR3 distances}
 \tablenotetext{c}{These may be young PNe rather than PPNe}
 \tablenotetext{d}{The high RUWE value for this PPN makes the distance much more uncertain.}
\end{table}

\subsection{Temperatures, Luminosities, and} Radii of PPN Central Stars \label{sec:PPNproperties}
The absolute physical properties of the central stars of PPNe have been somewhat elusive in the past. This is largely due to poorly known distances. While the spectra are well classified as supergiants (ranging from spectral types between G and B) these objects are clearly under-luminous compared to true high-mass evolved stars. And the spectral types, without precise distances, don't yield values for the absolute physical parameters. In Table \ref{tab:ppn_ages} we provide temperature and luminosity values from the literature for those objects where such values exist and are not based largely on assumptions of the evolutionary state of the object.

The recent availability of Gaia parallaxes has allowed reliable values to be determined for a number of post-AGB objects. \citet{oudmaijer2022} use Gaia DR3 data for 249 post-AGB candidate objects from a sample by \citet{vickers2015} for which the latter had calculated the total integrated flux based on fitting the spectral energy distribution, including correction for interstellar extinction.  For the stars in the sample with spectral types, \citet{oudmaijer2022} then assigned temperatures based on these. They find typical luminosities in the $10^3$--$10^4$ L$_\odot$ range with typical temperatures of $5000\lesssim$T$_\mathrm{eff}\lesssim25000$ K. Nearly one-third of their sample are, however, under-luminous for typical post-AGB evolutionary tracks (L$\lesssim1000$ L$_\odot$), which they claim as evidence of their post-RGB nature. We ignore these objects for our purposes. In addition, nine of the objects from \citet{oudmaijer2022} have spectral types of Wolf-Rayet stars. Due to complications with the interpretation of these objects, and because the central stars of PPNe are typically not WR stars, we also remove these objects from our present study. What remains are 119 post-AGB objects for which we use the \citet{oudmaijer2022} luminosity and T$_\mathrm{eff}$ values to calculate approximate radii (assuming blackbodies). The resulting radii, along with the spectral types, temperatures, and luminosities are provided in Table \ref{tab:postAGB_radii}. We find values $1.0\lesssim$R$\lesssim600$ R$_\odot$. Of the sample, 91 have radii $\gtrsim 10$ R$_\odot$. The remaining 28 have radii $\lesssim 10$ R$_\odot$, with ten of those having R$\lesssim 3$ R$_\odot$. Only seven have radii $>200$ R$_\odot$ with three $>400$ R$_\odot$ and one at $\sim400$ R$_\odot$. Some of these large and luminous objects may be massive evolved stars mis-classified as post-AGB stars. 
Apart from the under-luminous objects, these luminosities, temperatures, and radii generally agree well with expected values for post-AGB stars theoretical evolution tracks \citep[e.g.,][]{M3B2016}. All of the objects with R$\lesssim10$ R$_\odot$ have spectral type B and are thus consistent, based on the evolutionary tracks, with being much more evolved (e.g. older) and thus would have been larger previously when they were cooler. These sizes are particularly important in that the linear momentum observed in PPNe seem to require not just a CE phase, but for the companion to spiral in to a separation of $\lesssim10$ R$_\odot$ \citep[c.f.][]{blackman2014}.

We also find excellent agreement in temperature and luminosity between the post-AGB objects in Table \ref{tab:postAGB_radii} and the PPNe in Table \ref{tab:ppn_ages}, further supporting the connection between them and our use below of post-AGB stars as proxies for PPN cores.

%Table 2 - Post-AGB radii
\begin{longtable}{llcccc}
    
	\caption{Radii of post-AGB stars calculated from \citet{oudmaijer2022}}
	\label{tab:postAGB_radii}\\

\hline
GaiaDR3               &  SIMBAD             &    SpType   &  $\log T_\mathrm{eff}$ &   Lum      &  Radius \\
                   ~   &          ~           &       ~      &  [K]      &   L$_\odot$     &  R$_\odot$ \\
\hline
\endfirsthead
\hline
GaiaDR3               &  SIMBAD             &    SpType   &  $\log T_\mathrm{eff}$  &   Lum      &  Radius \\
\hline
\endhead

1369896865785991424            &     BD+33 2642              &    O7p           &           4.57    &           2260    &     1.1\\
5254793942926363392            &     IRAS 10214-6017         &    O7            &           4.57    &           4242    &     1.6\\
2032364166432389760            &     IRAS 19343+2926         &    B0            &           4.49    &           2156    &     1.6\\
6879196723703009920            &     IRAS 19590-1249         &    B1Ibe         &           4.42    &           1748    &     2.0\\
4124125282361429504            &     IRAS 17381-1616         &    B1Ibe         &           4.42    &           2984    &     2.6\\
6736747708089687936            &     IRAS 18371-3159         &    B1Iabe        &           4.42    &           3213    &     2.7\\
4099619470274753408            &     IRAS 18379-1707         &    B1IIIep       &           4.42    &           3268    &     2.7\\
6063703586653222144            &     IRAS 13266-5551         &    B1Iae         &           4.42    &           3560    &     2.9\\
4579182637944779264            &     IRAS 18062+2410         &    B1IIIe        &           4.42    &           3643    &     2.9\\
1328057763997734144            &     Cl* NGC 6205BARN29      &    B2p           &           4.36    &           2186    &     3.0\\
2005246464463628800            &     IRAS 22023+5249         &    B1            &           4.42    &           4264    &     3.2\\
 565507868441719424            &     IRAS 01005+7910         &    B2            &           4.36    &           2471    &     3.2\\
1869422453048750336            &     IRAS 20462+3416         &    B1            &           4.42    &           4411    &     3.2\\
4318134628803970816            &     IRAS 19306+1407         &    B0-1I         &           4.49    &           8501    &     3.2\\
2351623413515105920            &     BPS CS22946-0005        &    B             &           4.26    &           1237    &     3.5\\
4568163710366782848            &     PG 1704+222             &    B3            &           4.26    &           1242    &     3.6\\
5934701559547878144            &     IRAS 16133-5151         &    B0            &           4.49    &          17400    &     4.6\\
4136944866387751552            &     IRAS 17203-1534         &    B1IIIpe       &           4.42    &           9191    &     4.6\\
4049379725984965120            &     LS 4825                 &    B1Ib          &           4.42    &           9532    &     4.7\\
5946845601071213696            &     IRAS 17311-4924         &    B1Iae         &           4.42    &           9960    &     4.8\\
6715619076008049792            &     LSE 148                 &    B5            &           4.18    &           1253    &     5.2\\
6435349718091211264            &     LSE 237                 &    B5            &           4.18    &           1553    &     5.7\\
4067343654354938112            &     IRAS 17516-2525         &    B0            &           4.49    &          28400    &     5.9\\
5849958457496943744            &     IRAS 14331-6435         &    B3Ie          &           4.26    &           4103    &     6.5\\
4365451214021224320            &     LS IV -04 1             &    B             &           4.26    &           4248    &     6.6\\
5954670408684703872            &     IRAS 17476-4446         &    B7Ie          &            4.1    &           1028    &     6.8\\
5335709477519159936            &     IRAS 11353-6037         &    B5Ie          &           4.18    &           2552    &     7.4\\
5975083327400970752            &     IRAS 17277-3506         &    B4            &           4.22    &           6798    &     10.\\
5963059480546004608            &     IRAS 16594-4656         &    B7            &            4.1    &           2599    &     10\\
1194929381434604800            &     IRAS F15240+1452        &    B9Iab:p       &           4.02    &           1435    &     11\\
4128590918794710272            &     IRAS 17074-1845         &    B5Ibe         &           4.18    &           6698    &     11\\
6385794694664872320            &     CD-68 2300              &    B8            &           4.06    &           2472    &     12\\
5863599857739835264            &     IRAS 13064-6103         &    B             &           4.26    &          15800    &     12\\
4128590918794710272            &     IRAS 17074-1845         &    B5Ibe         &           4.18    &           8179    &     13\\
2020388045260203008            &     IRAS 19374+2359         &    B4            &           4.22    &          15000    &     14\\
2096072103492979584            &     V534 Lyr                &    A0Iabe        &           3.99    &           1930    &     15\\
1932229409071269248            &     BD+39 4926              &    B8            &           4.06    &           4050    &     16\\
3920735495441657728            &     BD+13 2491              &    B9            &           4.02    &           2855    &     16\\
6083719439934104832            &     CD-46 8644              &    A7            &            3.9    &           1245    &     18\\
2049984454412871296            &     IRAS 19200+3457         &    B8            &           4.06    &           7231    &     21\\
5237007177683569536            &     IRAS 11201-6545         &    A3Ie          &           3.95    &           2744    &     22\\
4120637086125583360            &     IRAS 17423-1755         &    B7e           &            4.1    &          11000    &     22\\
4686479751449676032            &     LB 3219                 &    B             &           4.26    &          48700    &     22\\
3159640386918214528            &     IRAS 07008+1050         &    A0            &           3.99    &           4068    &     22\\
5972489407656030720            &     IRAS 17208-3859         &    A2I           &           3.96    &           3171    &     22\\
2049034819957965312            &     IRAS 19410+3733         &    F3Ib          &           3.83    &           1016    &     23\\
5932016212933920384            &     LS 3593                 &    A0Ib          &           3.99    &           4453    &     23\\
5335675087769798272            &     IRAS 11387-6113         &    A3Ie          &           3.95    &           3254    &     23\\
2060616220769708672            &     IRAS 20145+3656         &    B             &           4.26    &          59700    &     24\\
5831295999979910656            &     IRAS 16206-5956         &    A0Iae         &           3.99    &           6024    &     27\\
4190636669164572928            &     IRAS 20023-1144         &    F2II          &           3.85    &           1763    &     27\\
 351149177434709760            &     IRAS 01427+4633         &    F2III         &           3.85    &           1913    &     29\\
5896479309853592448            &     IRAS 14072-5446         &    A3I           &           3.95    &           4860    &     29\\
5893945588395282304            &     IRAS 14488-5405         &    A0Ie          &           3.99    &           7420    &     30\\
 255225480926107392            &     IRAS 05040+4820         &    A4Ia          &          3.935    &           5255    &     32\\
5343168568718268800            &     IRAS 11385-5517         &    B8            &           4.06    &          17400    &     33\\
5903310335089068416            &     IRAS 15039-4806         &    A5Iab         &           3.92    &           5202    &     34\\
4046476534251259904            &     CD-30 15602             &    G0:           &           3.76    &           1217    &     35\\
3105987960396950784            &     IRAS 06530-0213         &    F5Ia          &           3.81    &           2298    &     38\\
6066902993687172608            &     IRAS 13110-5425         &    F5Ia/ab       &           3.81    &           2506    &     40\\
5462428643590805248            &     IRAS 10158-2844         &    B9            &           4.02    &          18000    &     40\\
6060828565581083264            &     IRAS 12360-5740         &    F0            &           3.87    &           4537    &     40\\
5351904394753372672            &     IRAS 10256-5628         &    F5I           &           3.81    &           2802    &     42\\
1958757291756223104            &     IRAS 22223+4327         &    F7I           &           3.79    &           2338    &     42\\
6070128028770373888            &     IRAS 13245-5036         &    A7            &            3.9    &           6863    &     43\\
3156171118495247360            &     IRAS 07134+1005         &    F5I           &           3.81    &           3296    &     45\\
4042544062195042176            &     IRAS 18023-3409         &    B9Ia+e        &           4.02    &          23200    &     46\\
1836195688380634368            &     IRAS 20160+2734         &    F3Ie          &           3.83    &           4076    &     46\\
2031794791233840128            &     IRAS 19454+2920         &    A0            &           3.99    &          18300    &     47\\
 994259335315643520            &     IRAS 06338+5333         &    F7IV:         &           3.79    &           3211    &     49\\
5707613169577769600            &     IRAS 08187-1905         &    F6Ib/II       &            3.8    &           3728    &     51\\
5869845594859880064            &     IRAS 13203-5917         &    G2I           &           3.73    &           2057    &     52\\
4035907203854415488            &     IRAS 18025-3906         &    G2I           &           3.73    &           2061    &     52\\
2020571869841643392            &     IRAS 19477+2401         &    F5I           &           3.81    &           4317    &     52\\
5520238967817034880            &     IRAS 08143-4406         &    F8I           &           3.78    &           3401    &     53\\
5849962851220246016            &     IRAS 14325-6428         &    F5I           &           3.81    &           4603    &     54\\
5515266327706463616            &     IRAS 08281-4850         &    F0I           &           3.87    &           8394    &     55\\
5906408788891928704            &     IRAS 14429-4539         &    G0Ie          &           3.76    &           3065    &     55\\
6871175064823382912            &     IRAS 19500-1709         &    F4Ia          &           3.82    &           5460    &     56\\
2074302426124470656            &     IRAS 19589+4020         &    F5I           &           3.81    &           5231    &     57\\
3497154104039422848            &     IRAS 12538-2611         &    F3Ia          &           3.83    &           6753    &     60\\
2902505745786910080            &     IRAS F05338-3051        &    G5            &            3.7    &           2050    &     60\\
1367102319545324288            &     IRAS 17436+5003         &    F3Ib          &           3.83    &           6978    &     61\\
4516723883521069952            &     IRAS 19207+2023         &    F6I           &            3.8    &           5655    &     63\\
4162959693758887424            &     IRAS 17279-1119         &    F2/3II        &           3.85    &           9313    &     64\\
4041945343757244160            &     IRAS 17440-3310         &    F3I           &           3.83    &           7975    &     65\\
3388902129107252992            &     IRAS 05113+1347         &    G5I           &            3.7    &           2615    &     67\\
4240112390324832384            &     IRAS 19386+0155         &    F5Ib          &           3.81    &           7253    &     68\\
3422437684728294528            &     IRAS 05140+2851         &    F0            &           3.87    &          14600    &     73\\
4061265519768800768            &     IRAS 17317-2743         &    F5I           &           3.81    &           8659    &     74\\
6076326701687231872            &     IRAS 12175-5338         &    A7I           &            3.9    &          20000    &     74\\
5617989266685365120            &     IRAS 07140-2321         &    F5            &           3.81    &           9548    &     78\\
 173086700992466688            &     IRAS 04296+3429         &    G0Ia          &           3.76    &           6168    &     79\\
4334241408966611328            &     IRAS 16476-1122         &    M1I           &           3.56    &           1013    &     80\\
6029425384023553792            &     IRAS 16559-2957         &    F5I(e)        &           3.81    &          10400    &     81\\
4582795323914832000            &     IRAS 17534+2603         &    F2Ibp         &           3.85    &          15300    &     82\\
2015785313459952128            &     IRAS 23304+6147         &    G2Ia          &           3.73    &           5242    &     83\\
5980714063986945920            &     IRAS 17106-3046         &    F5I           &           3.81    &          12800    &     90\\
4580154606223711872            &     IRAS 18095+2704         &    F3Ib          &           3.83    &          16600    &     94\\
5241806275407841664            &     IRAS 11000-6153         &    F2III         &           3.85    &          23700    &     102\\
2006425553228658816            &     IRAS 22272+5435         &    G5Ia          &            3.7    &           6166    &     104\\
4351018375858237952            &     IRAS F16277-0724        &    A7Ib          &            3.9    &          40700    &     106\\
2033763428091006720            &     IRAS 19475+3119         &    F3Ibe         &           3.83    &          22300    &     109\\
4158154754919296000            &     IRAS 18075-0924         &    G2I           &           3.73    &           9700    &     113\\
2034134414507432064            &     IRAS 20000+3239         &    G2I           &           3.73    &          10300    &     117\\
4072427555640528000            &     IRAS 18384-2800         &    F2/3Ia        &           3.85    &          37300    &     128\\
5824126771840487936            &     IRAS 15210-6554         &    K2I           &           3.62    &           4563    &     129\\
5864661779824561664            &     IRAS 13416-6243         &    G1I           &          3.745    &          14600    &     130\\
6130448958959242240            &     IRAS 12222-4652         &    F4            &           3.82    &          38000    &     149\\
 513671461473684352            &     IRAS Z02229+6208        &    K0            &           3.64    &           7978    &     156\\
4318934003785783680            &     IRAS 19396+1637         &    M7            &           3.47    &           1930    &     168\\
5698817012142459136            &     IRAS 08005-2356         &    F5Iae         &           3.81    &          54200    &     186\\
5853777267581362176            &     IRAS 14103-6311         &    M0            &           3.57    &           9921    &     240\\
5351069693654349952            &     IRAS 10456-5712         &    M0            &           3.57    &          16800    &     313\\
2030200671149815424            &     IRAS 20004+2955         &    G7Ia          &           3.67    &          52200    &     348\\
3303343395568710016            &     IRAS 03507+1115         &    M7            &           3.47    &          10700    &     396\\
5835411094089870592            &     IRAS 16099-5651         &    M8            &           3.47    &          12400    &     426\\
4264026012336768000            &     IRAS 19114+0002         &    G2Ia          &           3.73    &         181000    &     492\\
5941189713256986752            &     IRAS 16279-4757         &    M3II          &           3.49    &          28200    &     586\\
		\hline

\end{longtable}

\subsection{Photometric variability of PPNe}
Another set of measurements that can be helpful in our analysis is the photometric variability and longer-term stability of the stellar remnants inside the PPNe--whether we can observe measurable changes to the structure of the stellar remnant on scales of years or decades. The earliest photometric and radial velocity studies of PPNe found essentially all to have photometric variability, typically multi-periodic with a range of 30-160 days \citep[and references therein]{hrivnak2024}, and several of the brighter ones were monitored and found to have radial velocity variations as well, with the same period as the photometric variations \citep{hrivnak2013,hrivnak2018}. One of the stated goals of these surveys was to search for potential evidence of binarity or pulsations. To date all studies have shown that the photometric variations are primarily due to pulsational instability in these stars. The pulsations are also stable over the history of the photometric monitoring, which extends from around a decade \citep{hrivnak2010,arkhipova2010,hrivnak2024} up to 25 years \citep{hrivnak2022}. In fact, a monotonic relationship has been found in these objects between the pulsation period and T$_\mathrm{eff}$, with the hotter PPN cores having shorter pulsation periods, which is what would be expected from evolution toward higher temperatures at approximately constant luminosities. However, there is additional long-period variability in a small fraction of these systems. \citet{hrivnak2024} show periods of 5.2, 6.9, and 8.2 yrs (IRAS 07253-2001, 08005-2356, and 17542-0603) that could be evidence of relatively wide binary companions, and the binary nature of IRAS 08005-2356 was confirmed based on radial velocity measurements by \citet{manick2021}.

In addition to long-term periodic variability, there is also evidence of \emph{evolutionary} changes to the pulsation period and temperature over the course of a decade or more in two of the warmer PPNe \citep{hrivnak2022}. These period changes in the two warm PPNe are consistent with the expected evolution in temperature for an evolving post-AGB star \citep[e.g.,][]{M3B2016}. However, even these changes are stable (e.g. periodic or slow changes in period). These results show that the stars at the centers of these PPNe are long-term stable objects over the several decades of observation.

\section{Common Envelope Evolution} \label{sec:CEE}

The term \emph{common envelope evolution} is used to describe a range of phenomena in which a star expands as a result of evolutionary changes, becoming large enough to engulf a nearby companion. The companion and core of the evolved star orbit each other inside the larger envelope. Typically, this requires the engulfing star to have a clear core-envelope boundary since without one the definition of a ``common envelope'' is a difficult distinction to make. However, many scenarios have been proposed that fall under the mantle of CE evolution but which do not necessarily include two cores orbiting inside a ``common'' envelope. Some scenarios result in stellar mergers, some in envelope stripping by a nearby companion \citep[such as the hypothesized scenario of grazing envelope evolution, GEE,][]{soker2015}. Others end in a traditional CE event in which the binary survives or a CE event in which the engulfed star is shredded and merges with the stellar core. Thorough reviews of CE evolution modeling are provided by \citet{ivanova2013}, \citet{Ivanova2020}, and \citet{ropke2023} which detail the amazing progress made in understanding these events and the significant difficulties that remain.

\subsection{Stages of CE Evolution\label{CEstages}}
For the purposes of our discussion, an important summary of our current understanding is provided by \citet{ropke2023} who denote three distinct stages in CE evolution. These are the (\emph{i}) pre-CE; (\emph{ii}) dynamical inspiral/plunge, and (\emph{iii}) slow spiral-in/post-plunge stages. The first stage does not particularly concern us--while there may be considerable mass loss in the pre-CE phase \citep[e.g.][]{macleod2018,reichardt2019}, for PPNe the envelope has already been mostly ejected so that if it experienced a CE phase, the CE is either in progress or has been completed. So, here we will focus on the dynamical inspiral/plunge (hereafter simply referred to as the plunge stage) and slow spiral-in/post-plunge stages (hereafter the post-plunge stage).

One thing that has become clear in CE evolution studies is that the plunge stage happens on a dynamical timescale, with the orbital separation shrinking from the AU scale to tens of solar radii or less in time ranges of months to approximately a decade \citep{ivanova2013,ropke2023}, depending on the initial conditions and exactly when the formal plunge stage is determined to both begin and end. Until fairly recently, the general thinking was that the envelope was ejected either mostly or completely at the end of the plunge stage, primarily because the ejection of the envelope would halt the plunge (at least the plunge on dynamical timescales) and result in a short-period ($<2$ d) binary central star. However, more recent work suggests that the plunge may stop due to a reduction in drag from the lifting of the envelope or from co-rotation of the inner portion of the CE with the binary \citep[e.g.][]{gagnier2025,Bhattacharyya2026}. However, this may not completely eject the envelope and the actual ejection \emph{may} take longer, with the final energetic push due to ionization energy from recombination in the cooling gas as it expands. However, understanding exactly when and how the final envelope ejection occurs is still one of the primary difficulties in CE evolution simulations. It may well occur at the end of the plunge stage after all.

The post-plunge stage likely occurs more along thermal timescales, which can be on the order of $10^4$ years. During this stage the binary often, though not always, appears to continue to reduce its separation, but modeling difficulties over this large range of timescales (from dynamical timescales of days for the binary and structures of the components to thermal timescales of the remaining envelope) make this the most uncertain of the CE stages. 
In one study, \citet{ivanova2013} describe a self-regulating state that may develop in which the envelope expands but is not fully ejected during the plunge. Frictional energy from the inspiral during the plunge is transported to the surface of a thin remaining envelope and radiated away. The radius eventually increases to over 1000 R$_\odot$ while the central binary cores, which spiraled in to a separation of about 10 R$_\odot$ during the plunge phase, then slowly decrease their separation to closer to 1 R$_\odot$ during the self-regulating phase. It is unclear if this self-regulated situation occurs at all, whether the remaining envelope might actually form a disk due to high angular momentum from fallback material, or what are the observable properties of the remaining envelope. 
A few years later \citet{ivanova2016} used 3D models to explore a self-regulated phase showing slow inward evolution of the orbit as well as further mass loss through shell-triggered ejection or recombination runaway outflow during the post-plunge. However, they did not find co-rotation of the CE with the inner binary at any stage. They show an envelope during the post-plunge that is a spherical but clumpy shell
\citep[see][for a treatment of these issues]{iaconi2019}.

More recent work by \citet{gagnier2025} and \citet{Bhattacharyya2026} did not find a self-regulating phase, but rather a co-rotation of the inner CE with the central binary. In both, the inner co-rotating portion of the CE can be long-lived in the post-plunge stage with the inner binary continuing a slow
inward spiral. These post-plunge stages lasted at least hundreds to thousands of orbits of the central binary (a total of years to decades). And both studies result in a remaining envelope around the inner binary during the post-plunge.
\citet{gagnier2025} describe the resulting envelope as resembling a contact binary. \citet{Bhattacharyya2026} show a rotating outer ellipsoidal envelope with an inner density enhancement with the appearance of a contact binary.

Other models, such as those by \citet{sand2020} shown in \citet{ropke2023} as their Figure 6, show a final binary separation in the tens of R$_\odot$. In fact, most CE simulations have had a very difficult time moving binaries to the small orbital separations seen in the observed sample ($<10$ R$_\odot$, see below). This is one of the reasons that the post-plunge may be important---to finish reducing the binary separation to the final observed values. And it seems clear that these post-plunge stages {\it can} last from
decades up to $10^4$ yr.

\section{Close Binary Central Stars of Planetary Nebulae in Context}

Over one hundred close binary CSPNe have been identified\footnote{See https://www.drdjones.net/bcspn/ for an up-to-date list} so far. Orbital periods range from 0.068--18.1 days, with 80\% having orbital periods of $<2$ days.
Fewer than twenty-five of the catalogued close binary CSPNe have published physical properties based on full binary modelling that uses both light and radial velocity curves. Of the modeled systems one-third are double degenerate systems, with a white dwarf (WD) companion to the CS. The remaining two-thirds have a main sequence (MS) companion, typically an M dwarf, though some K dwarf and one G dwarf companions have been identified. Using light curve morphologies to predict companion types we find that of the full sample, approximately one-quarter of the close binary CSPNe have a WD companion and three-quarters of companions are MS stars.
No clear relationship has been identified between type of companion to the CS and the morphology of the surrounding PNe. %Does this track with CE models? Is there reason to believe that the size of the companion changes CE evolution?
So here we will operate under the assumption that companion type does not significantly directly affect the ejection of the CE (though observationally there {\it may} be a lower mass cutoff for companions that will give rise to the ejection).

Close binary CSPNe are especially important in helping to understand CE evolution because the PN \emph{is} the ejected envelope. In addition, PNe are typically visible for less than 50,000 years. Based on the relatively very short evolutionary times since ejection of the envelope, much too short for gravitational radiation or magnetic braking to have a significant effect, the current orbital separation of close binary CSPNe must be essentially equal to the separation at the end of the CE stage.
Based on the published binary models, the orbital separations range from 0.9--9.4 R$_\odot$. Since most have orbital periods less than a few days, the separations are heavily weighted to $<3-4$ R$_\odot$.

\section{Comparison of Properties}

By weighing the considerations above we can begin to evaluate whether PPNe, particularly those showing some bipolar or multi-polar structures and at least some elliptical nebulae, are the progenitors of close binary CSPNe. The primary properties that help us at this point are the sizes and ages of PPNe compared to the binary separations of close binary CSPNe and to the timescales of CE evolution.

If we compare spatial scales of each set of objects we find the following: the typical radii of the stellar remnant in PPNe are in the range 1--100 R$_\odot$, with the majority $>10$ R$_\odot$ \citep{blackman2014}, while close binary CSPNe have typical separations of 1--9 R$_\odot$. The plunge phase of CE evolution, based on models, tends to end with separations in the tens of R$_\odot$ or less, and in order to produce the linear momentum values observed in PPNe, the companion in CE evolution needs to spiral in to $<10$ R$_\odot$. So, the typical PPN central star is larger than the orbital separations of observed close binary CSPNe and is also larger than the typical core separations found by CE evolution after the plunge phase. Therefore, {\it if PPNe later evolve into close binary CSPNe}, we are left with one of the following possibilities for PPNe:
\begin{enumerate}
    \item the binary is not yet at its final separation and is in the CE plunge phase
    \item the binary is currently at, or near, its final separation and what we see as the CS of the PPN is in the post-plunge phase, but not changing orbital separation by more than a factor of 2--3.
    \item the binary is not yet at its final separation (the PPN CS has a nearby companion) but is in a post-plunge phase while the binary continues significant orbital separation reduction.
\end{enumerate}
We will address each of these possibilities further below.  

Option (1) above is effectively ruled out when we take into consideration the timescales involved. As we discussed above, the plunge phase of CE evolution occurs on dynamical times, meaning that the plunge phase for these stars will last at most years to decades. If the observed, resolved nebulosity in PPNe is the ejected CE, then the ages of even the youngest PPNe (see Table \ref{tab:ppn_ages}) are well beyond the plunge phase. The only way this would not be the case is if the ages in Table \ref{tab:ppn_ages} are in some way dramatically overestimated---they would need to be so by \emph{at minimum} a factor of 5 to 100, even assuming the longest plunge times from CE models. This would require incorrect distances (unlikely at that level given that many of these now have Gaia distances), or incorrect expansion models. While the uncertainties on expansion velocities and geometric models could possibly be large enough to move some of the PPN ages to small enough values, the corrections would need to move \emph{all} of the ages to much smaller values. There is currently no reason to believe that such a large systematic change in the modeling would be realistic.
Also, the nebulae in these PPNe are actually \emph{detached} from the remaining core, suggesting that the CE was ejected at a time in the past equal to the ages of the nebulosity. Additionally, a large sample of PPNe have been monitored photometrically and a smaller sample spectroscopically (for radial velocities) over several decades--the approximate maximum time for the plunge phase--with no apparent significant change in evolutionary state.

Option (2) is more difficult to address due to the lack of consistent models of the post-plunge phase as well as a lack of description of how this phase would be observed---in particular what observational characteristics it would display. Recent modeling seems to be more consistent, and some are producing models that can predict at least some observables \citep[see e.g.][]{gagnier2025}. The timescales for the post-plunge stage are of the same order of magnitude as those for PPNe---the thermal timescales of up to $10^4$ yr are consistent with the observed ages of PPNe. However, given the detached nature of the PPN nebulosity, the majority of the CE would need to have been ejected by the end of the plunge phase or at some point during the post-plunge, leaving a relatively low-mass remaining envelope that would need to resemble a stable stellar photosphere. Based on existing models we cannot rule this out, though it does dramatically limit the possible parameters of post-plunge evolution. For example, PPNe have temperatures, spectra, and radii that are consistent with observed post-AGB stars (see \S\ref{sec:PPNproperties}). For those PPNe that have been monitored, some for two decades, those values have been incredibly consistent, with some evidence of changes consistent with post-AGB evolutionary models \citep{hrivnak2022,M3B2016}. Furthermore, most of the PPNe with photometric monitoring show variability due to radial pulsations. Therefore, for option 2 to hold, any post-plunge models would need to reproduce an envelope that looks like a stable stellar atmosphere consistent with observed post-AGB stars and calculated models, and an envelope that is unstable to radial pulsations. We recommend that future modeling efforts attempt to produce predictions of \emph{observables} for their post-plunge systems. For recent papers that have described some observables, those are inconsistent with observed PPNe. \citet{gagnier2025}, for example describe a stable rotating system, but one that resembles a contact binary--something that would show up clearly in both photometric and radial velocity studies but we do not see in PPNe. The ellipsoidal envelope in \citet{Bhattacharyya2026} suffers from the same issues as a potential explanation for PPNe. Therefore, unless future models can produce a decades-stable post-AGB-like pulsating stellar photosphere it seems at unlikely that option 2 could describe the observed PPNe.

Note that we have characterized option (2) as the binary being at, or near, the final separation, which limits the post-plunge evolution even more dramatically. In this case all, or nearly all, of the actual orbital separation reduction would have occurred before this stage. However, as we note above, some continued orbital spiral-in may still occur in the post-plunge phase (as discussed in \S\ref{CEstages}). Thus, for option (3) we consider the case in which further orbital evolution occurs during the PPN stage. Here the PPN core and companion star are separated, but continue to spiral in as the PPN core shrinks over time, eventually arriving at the small orbital separations we observe in close binary CSPNe. Note that this means that what we currently observe as the PPN core is {\it only} the core of the original star and would \emph{not} be a remaining envelope surrounding a close binary as in the post-plunge models \citep[e.g.][]{ivanova2016,gagnier2025,Bhattacharyya2026}. The companion would exist outside of the observed PPN core. Most CE models (see \S\ref{CEstages}) only show orbital separation reductions of at most an order of two, so already this option is not supported by most models. But there are additional difficulties with this option. First, given the sizes of the PPN core remnants, \emph{all} of the observed PPNe would have had the plunge stop at separations greater than about twice their current PPN core radius (depending of course on the mass of the companion, but here we are making a qualitative argument). If the separation were lower, then the companion would either be accreting mass from the PPN core or would at least partially deform the PPN core towards its Roche lobe resulting in ellipsoidal variability during the orbit, which we do not observe. Second, from studies of radial velocity \citep{hrivnak2017} and long-term optical photometry \citep{hrivnak2022,hrivnak2024}, very few PPNe appear to be in binary systems with these types of intermediate orbital periods (1--few years) making this
option unlikely.

Finally, for intermediate period binaries in PPNe to continue to spiral in to closer separations would require some type of drag. It is not clear from observations of PPNe what could cause this drag in the region immediately around the PPN core. Although some mass loss is expected, and perhaps observed \citep{hrivnak2024}, it would likely not create enough drag to significantly reduce the orbital period and would more likely result in mass loss from the system that actually \emph{expands} the orbit. 
For these reasons we also find option (iii) to be unlikely.

\section{Conclusions} \label{sec:Conclusions}

We find that each of the options listed above by which a PPN would evolve through a CE phase into a close binary CSPN is unlikely. Unless there exists a stable stage of post-plunge in which the evolved star retains a small fraction of the overall envelope that can appear as a stable post-AGB-like pulsating photosphere for at least several decades, PPNe are most likely \emph{not} the progenitors of close binary CSPNe.

This then raises the question of {\it where do the identified PPNe candidates fall in the evolution from AGB stars to the central stars of PNe}? One possibility, which maintains the connection with CE systems, is that these experienced a CE phase in which the companion was either disrupted or merged with the core. In these cases, the plunge would have contributed energy to the envelope expansion and could have helped with the shaping of the ejected envelope, but perhaps would not have resulted in the complete ejection of the envelope. This may occur if the envelope was lifted enough prior to disruption/merger to then allow recombination energy to complete the ejection, through dust-driven winds \citep{Glanz2018}, or the ejection may have completed via normal AGB mass loss - possibly with some enhancement. In effect it would be similar to the CE modeling in which the plunge ends due to reduced drag, perhaps due to corotation of the inner envelope with the core and companion, but in this case without a remaining companion. 

The idea that PPNe may be disruption/merger events has been discussed previously \citep[c.f.][]{kaminski2018} and we find it here to be a promising scenario for the formation of strongly shaped PPNe based on the current state of our understanding.

Another possible evolutionary scenario for PPNe is that they are binary systems which did not enter a formal CE phase, but experienced Roche lobe overflow, possibly wind Roche lobe overflow \citep{mohamed2007}. This could result in the formation of jets which aid in the ejection and shaping of the envelope. One example of this may be grazing envelope evolution mentioned above \citep{soker2015}. But again, few appear to have intermediate period companions of a year to decades. So either the companion is very wide, which would reduce the likelihood of interaction, or the companion was disrupted in the interactions.
The main argument for binarity in PPNe is based on the non-spherical, and in a significant number of cases, bipolar or multipolar shapes of their circumstellar envelopes.  
Recent high spatial resolution submillimeter wavelength observations with ALMA show that the bipolar and ring/spiral structures seen in PNe are already present in AGB stars.  Modeling of these images indicates that these structures can be produced by long-period stellar or substellar companions \citep{decin2020}. 

A third evolutionary scenario is that PPNe are effectively single stars that are evolving from the AGB to PN phases.
This was the initial explanation, since they have the spectra of G$-$B supergiants, are surrounded by cool dust and expanding molecular envelopes, and in some cases show evidence of third dredge-up elements in their spectra \citep{kwok1993,vanwinckel2003}.
A long-term radial velocity monitoring study of seven of the brightest PPNe showed little evidence for binarity \citep{hrivnak2011,hrivnak2017}, indicating that any binary companions must be long-period ($\geq$ 30 yrs) or low-mass ($\leq$ 0.2 M$_{\sun}$). Overall, only a few definite or likely cases of binarity have been found in PPNe, and none confirmed to have a short period ($\leq$ 1 yr)  \citep{sanchezcontreras2004,manick2021,hrivnak2024}. However, given our discussion in \S\ref{PPN shaping} this does not appear to be a viable option for producing the strongly shaped and highly energetic PPNe, although it may be able to produce the spherical and elliptical PPNe that do not possess high-momentum outflows, so at present we reject it as a suitable option.

Since it appears that PPNe are not the precursors of close binaries in CSPNe, another question emerges: {\it where are the precursors of the close binaries}? They must go through a similar process as described above---that is, some CE phase and shaping of the ejecta. Have we not observed them, or have they been seen but not yet identified properly? It seems likely that a close binary at the observed separations effectively bypasses the PPN phase. The objects we classify as PPNe are cases in which the central star is not yet hot enough to ionize the surrounding gas. In surviving close binaries, such a close plunge would likely force the hot core to evolve faster (nearly instantaneously due to removal of the outer layers) to 30,000 K or hotter such that it almost immediately ionizes the surrounding nebula. This seems to be the most likely scenario. Another possibility, that we believe is less likely, is that some PPNe are close binaries surrounded by enough remaining material that the central binary is enshrouded and not observed. In which case, these should be detectable as a dust-enshrouded IR source.
% Add discussion of specific bianry CSPNe to CSPN section above? Especially young examples (NGC 6026, Abell 41, Hen 2-428) and discuss their evolutionary state as well as their nebulae relative to PPNe?

It is possible that our understanding of these systems--the PPNe and the PNe with close binary central stars--will rely on continuing advances in CE evolution modeling. Although such modeling has made enormous strides in the past few decades, it does not yet provide a complete picture. Certainly, our understanding of these nebulae will depend on a better understanding of the ejection and shaping of the circumstellar envelope by a low mass stellar (or planetary) companion.

\begin{acknowledgments}

This material is based upon work supported by the U.S. National Science Foundation under award No. AST2107768. Any opinions, findings and conclusions or recommendations expressed in this material are those of the author(s) and do not necessarily reflect the views of the U.S. National Science Foundation.

DJ acknowledges support from the Agencia Estatal de Investigaci\'on del Ministerio de Ciencia, Innovaci\'on y Universidades (MCIU/AEI) and the European Regional Development Fund (ERDF) under grants PID2022-136653NA-I00 and PID2025-172507NB-I00 (DOI:10.13039/501100011033). DJ also acknowledges support from the Agencia Estatal de Investigaci\'on del Ministerio de Ciencia, Innovaci\'on y Universidades (MCIU/AEI) and the the European Union NextGenerationEU/PRTR with reference CNS2023-143910 (DOI:10.13039/501100011033). DJ also acknowledges financial support from the Agencia Estatal de Investigaci\'on (AEI) through the Severo Ochoa Centre of Excellence accreditation awarded to the Instituto de Astrof\'isica de Canarias, grant CEX2025-001609-S, funded by MICIU/AEI/10.13039/501100011033.

This work has made use of data from the European Space Agency (ESA) mission Gaia (https://www.cosmos.esa.int/gaia), processed by the Gaia Data Processing and Analysis Consortium (DPAC, https://www.cosmos.esa.int/web/gaia/dpac/consortium). Funding for the DPAC has been provided by national institutions, in particular the institutions participating in the Gaia Multilateral Agreement.

\end{acknowledgments}

\vspace{5mm}
\facilities{}

\software{}

\bibliography{PPNe-bCSPNe}{}
\bibliographystyle{aasjournal}

\end{document}